Title: Practical challenges in mediation analysis: A guide for applied researchers

Authors: Megan S. Schuler,[a] Donna L. Coffman,[b] Elizabeth A. Stuart,[c,d,e] Trang Q. Nguyen,[d] Brian Vegetabile,[f] Daniel F. McCaffrey[g]

a)  RAND Corporation, Arlington VA
b)  Department of Psychology, University of South Carolina, Columbia SC
c)  Department of Biostatistics, Johns Hopkins Bloomberg School of Public Health, Baltimore MD
d)  Department of Mental Health, Johns Hopkins Bloomberg School of Public Health, Baltimore MD
e)  Department of Health Policy & Management, Johns Hopkins Bloomberg School of Public Health, Baltimore MD
f)  RAND Corporation, Santa Monica CA
g)  ETS, Princeton, NJ

STATEMENTS AND DECLARATIONS:

No authors reported any financial or other conflicts of interest in relation to the work described.

**Funding**: This work was supported by funding from grants 1R01DA034065 and P50DA046351 from the National Institute on Drug Abuse and R01MH115487 from the National Institute of Mental Health. The content is solely the responsibility of the authors and does not necessarily represent the official views of NIDA, NIMH, the NIH or the US Government.

ABSTRACT

Mediation analysis is a statistical approach that can provide insights regarding the intermediary processes by which an intervention or exposure affects a given outcome. Mediation analyses rose to prominence, particularly in social science research, with the publication of Baron and Kenny's seminal paper and is now commonly applied in many research disciplines, including health services research. Despite the growth in popularity, applied researchers may still encounter challenges in terms of conducting mediation analyses in practice. In this paper, we provide an overview of conceptual and methodological challenges that researchers face when conducting mediation analyses. Specifically, we discuss the following key challenges: (1) Conceptually differentiating mediators from other "third variables," (2) Extending beyond the single mediator context, (3) Identifying appropriate datasets in which measurement and temporal ordering supports the hypothesized mediation model, (4) Selecting mediation effects that reflect the scientific question of interest, (5) Assessing the validity of underlying assumptions of no omitted confounders, (6) Addressing measurement error regarding the mediator, and (7) Clearly reporting results from mediation analyses. We discuss each challenge and highlight ways in which the applied researcher can approach these challenges.

# INTRODUCTION

Mediation analysis is a statistical approach that can provide insights regarding *how* an intervention or exposure affects a given outcome by investigating potential intermediate variables (aka "mediators") that are influenced by the exposure and, in turn, influence the outcome. Specifically, for an *a priori* proposed mediator, mediation analysis can evaluate to what extent the effect of the exposure on the outcome occurs through a mechanistic pathway involving the mediator compared to all other pathways that do not involve the mediator.

Mediation analysis has primarily been applied in the fields of health and social science research to investigate potential mediating pathways in the context of both randomized interventions (e.g., medication regimen or parenting skill-building program) or non-randomized exposures (e.g., exercise self-efficacy or COVID-19 related school closures). Many behavioral interventions (e.g., nutrition education program) are designed to impact the outcome of interest (e.g., cardiovascular health) by addressing an intermediate factor (e.g., diet). Mediation analysis can provide insights as to what extent the intervention impacted the outcome through the hypothesized mediating pathway(s), which can facilitate design of future interventions that are more effective and/or more parsimonious. Some agencies funding health and social science research – e.g., the US National Institute of Mental Health (NIMH) – now encourage the use of mediation analysis for investigating mechanisms of intervention effects (Cashin et al., 2019; Nguyen et al., 2020). In observational studies, mediation analyses can help to identify novel mediating pathways linking a non-randomized exposure and outcome (e.g., through what processes did the COVID-19 pandemic impact depression?), which may be targeted by subsequent intervention development.

We note that although mediation analysis has seen an increased application in the field of program evaluation, it is rarely applied, to date, in the context of policy analysis (Keele et al., 2015; Ludwig et al., 2011). However, mediation analysis can similarly offer insights as to the means through which local, state, or federal policies may impact population outcomes. Having a mechanistic understanding of how certain policies work can help policymakers to focus resources on the key elements of the policy that are driving outcomes as well as to consider alternative policy strategies if policies are found to have minimal effect on the desired intermediary pathways.

Despite the growing popularity of mediation analysis methods, applied health services and health policy researchers may still encounter challenges in terms of implementing such analyses in practice. In this paper, we review 7 key challenges, ranging from conceptualizing scientific questions regarding mediation to analytic challenges, such as how to address measurement error in mediation analyses. We hope this review provides pragmatic guidance for researchers implementing mediation analyses.

## BACKGROUND ON MEDIATION ANALYSIS

**Figure 1** depicts the traditional mediator model, linking exposure *A,* mediator *M,* and outcome *Y*. Specifically, all relationships depicted in the mediation conceptual model – namely, the effect of *A* on *M*, the effect of *M* on *Y*, and the effect of *A* on *Y* – are hypothesized to be causal. There are two possible causal paths from *A* to *Y*: (1) the path from *A* to *Y* that goes through the mediator *M* (referred to as the *indirect effect* of *A* on *Y*) and (2) the path from *A* to *Y* that does not go through *M* (referred to as the *direct effect)*. The indirect effect is generally of primary interest, as this represents the mediated pathway. The direct effect quantifies the relative magnitude of the effect of *A* on *Y*, through all other pathways that do not involve the mediator.

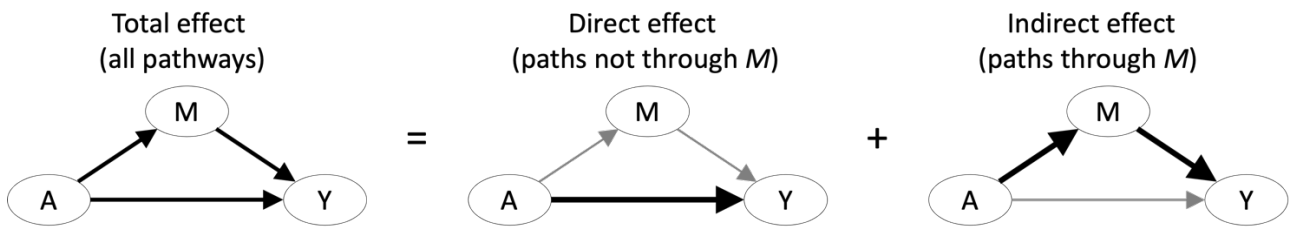

**Figure 1.** Mediation: Decomposition of the total effect into the direct effect and indirect effect

### Traditional approach:

Mediation analyses rose to prominence, particularly in social science research, with the publication of Baron and Kenny's (Baron & Kenny, 1986) seminal paper which has been cited nearly 125,000 times to date. Traditional mediation analysis required a continuous mediator and outcome and was estimated using linear regression. Under the parametric models specified in the traditional mediation framework the sum of the indirect effect and the direct effect exactly equals the *total effect* of *A* on *Y*. Baron and Kenny originally proposed a 3-step approach comprised of: (1) confirming the relationship between the exposure and outcome by regressing the outcome on the exposure, (2) confirming the relationship between the exposure and the mediator by regressing the mediator on the exposure and (3) confirming the relationship between the mediator and the outcome by regressing the outcome on both the exposure and the mediator. Conceptually, the coefficient on the mediator would be significant and the coefficient on the exposure would be attenuated relative to the coefficient estimated in Step 1. The estimate of the total effect is the exposure coefficient from the Step 1 model (outcome ~ exposure) and the estimate of the direct effect is the exposure coefficient from the Step 3 model (outcome ~ exposure + mediator). The estimate of the indirect effect is calculated as the difference between the estimates for total effect and the direct effect. Additionally, Sobel's test can be used to test for the presence of a significant indirect effect, namely presence of significant mediation. We note that it has now been recognized that mediation can exist without Step 1 being satisfied ( i.e., mediation in the absence of a significant total effect) (O'Rourke & MacKinnon, 2015).

**Modern causal inference approach:**

Methodological work subsequently highlighted certain key limitations of the traditional approach to mediation analysis, relating both to the analytic approach as well as the study design (e.g., lack of attention to confounding and temporality). A major advancement in more recent years is the incorporation of a causal inference framework and methodology (Imai et al., 2010; Pearl, 2001; Robins & Greenland, 1992; VanderWeele, 2015). The causal mediation analysis framework distinguishes between three distinct steps, (1) causal effect definitions, (2) causal effect identification, and (3) causal effect estimation. In this framework, causal effects are defined as the difference between two potential outcomes (rather than based on quantities from parametric models) (Holland, 1986). There are several different types of effect definitions (to be introduced under Challenge 4 later in the paper), which are more general than those permitted by the traditional approach. Resultantly, mediation analyses can be conducted in settings with binary variables and allowing exposure-mediator interactions (Hong et al., 2015; Moerkerke et al., 2015; Rijnhart, Valente, et al., 2021; Valeri & Vanderweele, 2013). Multiple recent papers have detailed the relationship between traditional mediation analysis and causal mediation analysis, including describing under which conditions and assumptions both approaches would yield the same estimates (MacKinnon et al., 2020; Rijnhart et al., 2017; Rijnhart et al., 2019; Rijnhart et al., 2023).

Estimation of the causal effects is a separate distinct step and can be conducted using either parametric or nonparametric models. Notably, causal mediation is a field of active methodological development, with many distinct estimation methods proposed for various causal mediation effects and particular recent growth in nonparametric and machine learning approaches. Broadly, estimation approaches primarily entail regression, weighting, or simulation (Nguyen et al., 2023). Estimation methods may differ not only with respect to which specific causal mediation effects they estimate, but also on factors such as bias/variance trade-offs and robustness to model misspecification. Additional considerations may include the nature of the outcome – e.g., alternative methods are needed in the context of a survival outcome measured as time-to-event variable (Fairchild et al., 2019; Tchetgen Tchetgen, 2011; VanderWeele, 2015; Vansteelandt et al., 2019; Vo et al., 2022). We note that a full discussion of the specific estimation approaches in causal mediation analysis is beyond the scope of this paper – please see (Nguyen et al., 2023) for a relatively accessible introduction to several common strategies used in effect estimation, as well as citations for many important papers on this topic. Finally, the causal inference framework has served to clarify the set of underlying assumptions required to estimate specific mediation effects.

**CHALLENGE 1: Conceptually differentiating mediators from other "third variables"**

Fundamentally, a mediation analysis must specify a presumed causal model that forms the basis for the proposed analyses. Indeed, mediation analysis properly falls under the umbrella of causal inference methods, as mediation analysis inherently examines causal effects (Coffman, 2015; Nguyen et al., 2020), a point consistently highlighted in the mediation literature (e.g., Baron & Kenny, 1986; MacKinnon, 2008; Preacher, 2015). That is, mediators are variables that lie in the causal pathway between exposure and outcome, and the mediator is understood to be part of a pathway that conveys (some of) the effect of the exposure to the outcome.

However, in practice, when one is constructing a causal model, it can be challenging to conceptually identify the mediating variable of interest. Generally, the exposure (treatment) and outcomes are readily identifiable, yet numerous "third variables" may be considered as potential mediators (Coffman, 2015). As we review here, a given third variable may represent a mediator, or it may more appropriately be conceptualized as a moderator, confounder, or collider. As we detail below, these 4 types of third variables – mediators, moderators, confounders, and colliders – are distinguished by their specific relationships to the exposure and the outcome (**Figure 2**).

**Figure 2**. Distinguishing third variables: mediators, moderators, confounders, and colliders

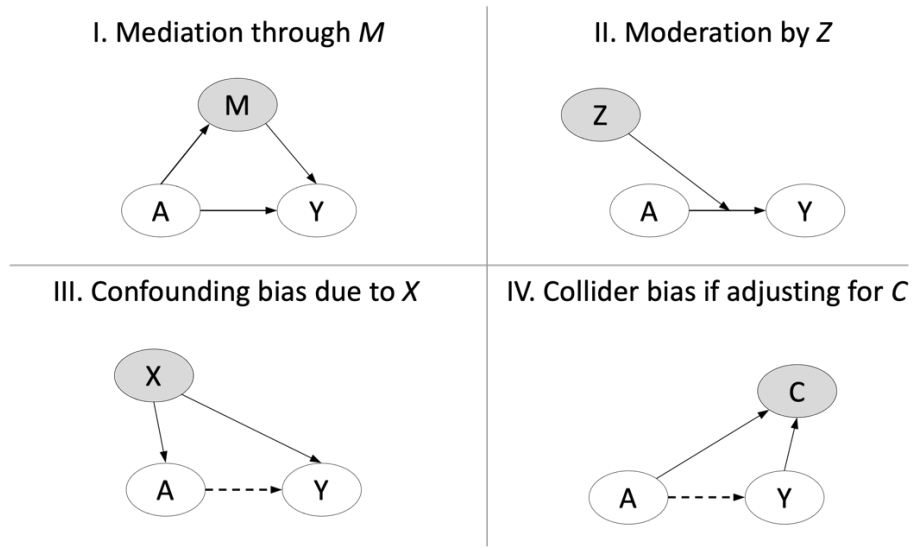

A mediator, commonly denoted *M*, is the only type of third variable that lies in the causal path between the exposure (denoted *A*) and outcome (denoted *Y*). Essentially, the relationships between *A, M,* and *Y* (i.e., the effect of *A* on *M*, the effect of *M* on *Y*, and the effect of *A* on *Y*) – are hypothesized to be causal. As an illustrative example, one mediator in the path between socioeconomic status (SES) and cardiovascular disease may be diet. An individual's diet may be directly influenced by their SES and may in turn impact their cardiovascular health.

A moderator (commonly denoted *Z)*, shown in Panel II of **Figure 2**, represents another type of third variable. A moderator is a variable that alters the effect of the exposure on the outcome, resulting in effect heterogeneity. Note that a moderator is not directly on the causal path between *A* and *Y*, but rather the level of the moderator variable determines the magnitude of the *A-Y* effect. That is, moderation questions seek to determine *for whom* and *under what conditions* a given exposure affects the outcome (Hong, 2015). For example, if a certain type of medication has a larger effect on blood pressure for women compared to men, then sex is a moderator of the medication–blood pressure effect. Moderating variables may include individual characteristics (e.g., sex may impact the pharmacological effects of a medication), contextual characteristics (e.g., a clinician's level of training may impact how effectively they provide psychotherapy), and prior or concurrent treatments (e.g., whether a patient previously received a specific type of chemotherapy may impact how effective current radiation therapy is). While some have asserted that moderators must temporally precede the exposure whose effect they moderate (e.g., (Kraemer et al., 2008)), Hong clarifies that moderators neither need to occur prior to the exposure nor be independent of the exposure (see (Hong, 2015) for a detailed discussion of moderated treatment effects).

An additional type of third variable is a confounder (commonly denoted *X*), as depicted in Panel III of **Figure 2**. A confounder represents a common cause of both *A* and *Y*. For example, in the context of investigating the effect of Alzheimer's disease on mortality, age would represent a confounder, as age is causally associated with both Alzheimer's disease and mortality. Comparing confounders and mediators, we note that a confounder both precedes and causes *A* whereas a mediator is caused by *A* and thus occurs subsequent to *A*. Note that the effect of *A* on *Y* is shown as a dashed line to highlight that a confounder can distort the true effect of *A* on *Y* – indeed, the presence of a confounder may induce the appearance of an effect between *A* and *Y* when none truly exists. To avoid confounding bias, confounding must be addressed on the design side via randomization or on the analytic side via statistical adjustment or stratification (Lipsky & Greenland, 2022).

A final type of third variable is a collider (denoted *C*) as shown in Panel IV of **Figure 2**. A collider variable is one that has two variables causally leading into it. We note that while the figure depicts *A* and *Y* as these 2 causal variables, definitionally it does not need to be *A* and *Y* (e.g., it could be *A* and *M*). In some sense a collider is the opposite of a confounder, as a collider is a common outcome of *A* and *Y* as opposed to a common cause. For example, consider an individual's socioeconomic status during childhood as the exposure of interest and their high school GPA as the outcome of interest. Both of these factors likely influence one's educational status at age 25, which would represent a collider in this context. Collider bias occurs when an observational study either disproportionately samples individuals based on the collider variable (e.g., only includes individuals with at least a college degree) or stratifies on the collider variable in the analysis – in both scenarios, the observed effect of *A* on *Y* will represent a distortion of the true effect of *A* on *Y*. To avoid collider bias, sampling or stratification based on the collider should be avoided (Holmberg & Andersen, 2022).

**Table 1.** Summary of different types of "third variables"

|  | *Definition:* | *Illustrative example:* |
| --- | --- | --- |
| Mediator | A mediator is the only type of third variable that lies in the causal path between the exposure and outcome. | One mediator in the path between socioeconomic status (SES) and cardiovascular disease may be diet. SES may directly impact one's diet, which may in turn impact their cardiovascular health. |
| Moderator | A moderator is a variable that alters the effect of the exposure on the outcome, resulting in effect heterogeneity. | If a certain type of medication has a larger effect on blood pressure for women compared to men, then sex is a moderator of the medication's effect. |
| Confounder | A confounder represents a common cause of both the exposure and outcome. | If age is causally associated with both Alzheimer's disease and mortality, then age would confound the relationship between Alzheimer's disease and mortality. |
| Collider | A collider variable is one that has two variables causally leading into it (i.e., it is a common outcome of two variables).. | If both an individual's childhood socioeconomic status and their high school GPA influence one's educational status at age 25, then age 25 education status represent a collider with respect to childhood SES and high school GPA. |

It is imperative to remember that statistical analyses generally cannot empirically differentiate whether a given third variable is "truly" one of the above four types. Rather, the applied researcher must *a priori* specify a third variable's role (based on their contextual knowledge) and then subsequently determine the most appropriate way to analytically model the variable. An additional point of confusion is that the same "variable" could represent either a moderator or a mediator in different contexts (Coffman, 2015). For example, consider the relationship between depressive symptoms and suicidality. Evidence suggests that social support acts as a moderator, in that higher levels of social support can attenuate the relationship between depressive symptoms and suicidality (Rubio et al., 2020). Social support is not on the hypothesized causal path between depressive symptoms and suicidality, and so is not a mediator in this context. In contrast, consider a community-building intervention for older adults that seeks to reduce loneliness by improving social support. In this context, level of social support is a hypothesized mediator in the causal path linking the intervention and loneliness. Thus, a given factor – e.g., social support – is not inherently a mediator or moderator; rather, the hypothesized causal relationships determine the role of a specific variable in a given analytic context. Furthermore, in the context of mediation, statistical analyses cannot provide direct evidence of causality. Analytic results provide insights regarding observed associations, given the hypothesized conceptual model. As highlighted by Holland (1986), the challenges of causal inference lie in inferring evidence of causality from these observed associations, which generally reflect a mix of causal effects and various non-causal components, and in assessing the validity of the assumptions required to interpret relationships as causal.

## CHALLENGE 2: Extending beyond the single mediator context

While mediation is often discussed with respect to a single mediator, in practice there are often multiple potential mediators.

*Illustrative example:* In an intervention context, consider a parenting intervention program that seeks to reduce substance use among parents of young children. The program specifically targets three constructs that may drive substance use behaviors: depression, parenting stress, and household poverty (Massarwi et al., 2021). When evaluating effectiveness of this intervention, examining the mediating roles of all three constructs may be of interest. The following figure depicts the relationship between the intervention program, the 3 mediators, and the outcome (assuming independence of the 3 mediators).

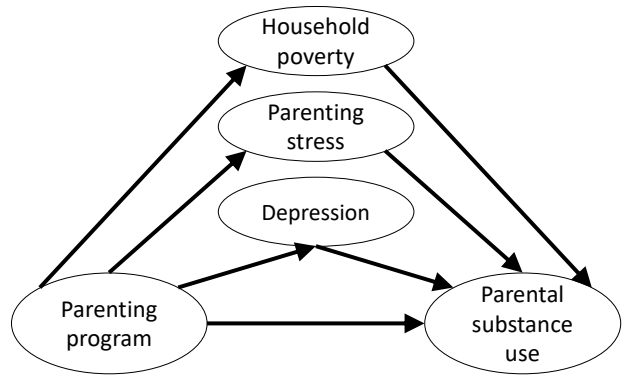

To date, the mediation literature has largely focused on applications with a single mediator, or the straightforward extension in which a set of mediators are treated as single *en bloc* composite variable, allowing application of traditional single mediator methods (Daniel et al., 2015). When multiple mediators are present, a host of conceptual and analytic questions arise: Are the mediators causally ordered, such that earlier mediators may affect later mediators? If the mediators are not causally ordered, are they assumed to be independent or allowed to be correlated? A growing literature on multiple mediator analysis has emerged, highlighting the complexities (and required assumptions) of identifying path-specific effects through multiple mediators. As shown in **Figure 3,** in the setting in which there are 2 causally ordered mediators, $M_1$ and $M_2$, there are 3 different indirect effects – one strictly through $M_1$, one strictly through $M_2$, and one through both. For references on the setting in which mediators causally affect each other see: (Daniel et al., 2015; Gao & Albert, 2019; Imai & Yamamoto, 2013; Steen et al., 2017; Tai et al., 2021); for the setting in which multiple mediators do not causally affect each other see: (Jerolon et al., 2020; Lange et al., 2013; MacKinnon, 2015; Preacher & Hayes, 2008; Taguri et al., 2018).

**Figure 3.** Conceptual model with 2 causally ordered mediators, $M_1$ and $M_2$

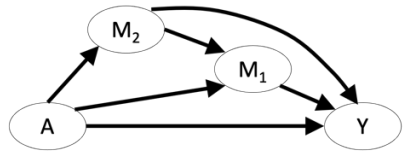

**3 different indirect effects through $M_1$ and/or $M_2$**

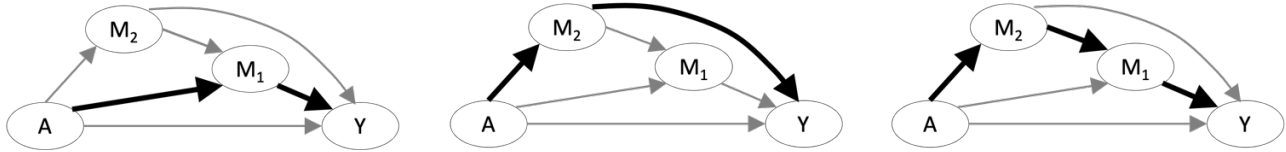

Alternatively, it may be the case that an additional intermediary variable – here denoted as *X* – is not of substantive interest as a mediator, but rather represents a potential post-treatment confounder (Coffman, 2015). Specifically, *X* occurs after the exposure *A* (and is influenced by *A*) and precedes *M*, functioning as a confounder of the mediator and outcome (see **Figure 4**). Notably, *X* is on the causal path between *A* and *Y*, and thus is a candidate mediator. As shown in **Figure 4,** the presence of *X* results in two indirect effects (both through and independent of *X*) as well as two direct effects (both through and independent of *X*). In this context, we can define additional effects of interest, namely the *partial indirect effect, total indirect effect, partial direct effect,* and *total direct effect* – each of which may be of interest depending on the scientific question under investigation. The *partial indirect effect* consists only of the pathway that goes through the mediator but not through the post-treatment confounder, whereas the *total indirect effect* consists of all pathways that go through *M*. Similarly, the *partial direct effect* consists only of the path between the exposure and the outcome that does not involve either the mediator or the post-treatment confounder, whereas the *total direct effect* consists of all pathways that do not go through *M*. It is essential to appropriately account for post-treatment confounding, as treating a post-treatment confounder like a baseline confounder (e.g., using regression adjustment) will result in biased estimates of the direct effect. See (Coffman & Zhong, 2012; De Stavola et al., 2015; Hong et al., 2022; Miles et al., 2020; Moerkerke et al., 2015; Valente et al., 2017) for more discussion.

**Figure 4.** Illustration of the additional direct and indirect paths between the exposure *A* and outcome *Y* in the presence of a post-treatment confounder *X*

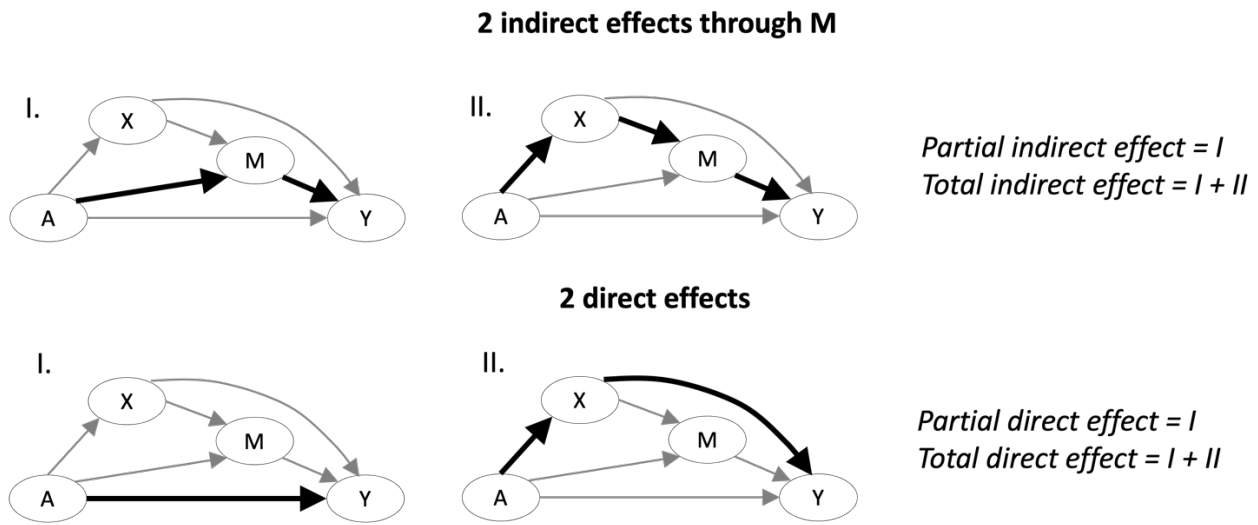

Another extension of the traditional single mediator model is the time-varying context, in which any of the exposure, mediator, outcome, and third variables may be dynamically changing across time. In this complex and challenging context, previous assessments of the time-varying exposures, mediators and outcomes act as post-treatment confounders or additional mediators. Multiple methods have been proposed for mediation analysis in this context, using approaches such as latent growth models, multilevel structural equation models, autoregressive models, and sequential mediation models. See (Gunzler et al., 2014), (VanderWeele & Tchetgen Tchetgen, 2017), (Lin, Young, Logan, et al., 2017), (Berli et al., 2021), and (Cai et al., 2022) for more details. This area remains an active field of methodological development.

*Illustrative example:* A pharmacological medication for smoking cessation may impact smoking outcomes by reducing cravings for cigarettes. However, the degree of cravings that a person experiences may decrease as the time since quitting smoking increases. Hence, craving may act as a time-varying mediator. See (Cai et al., 2022) for further examples.

**CHALLENGE 3: Identifying appropriate datasets in which measurement and temporal ordering supports the hypothesized mediation model**

As originally included in Hill's criteria, temporality is a core aspect of causality (Hill, 1965), in that the treatment or exposure of interest must come before the outcome in order to be deemed a causal factor – i.e., "cause precedes effect." In the mediation context, this temporal ordering requirement applies to multiple paths, as the treatment should temporally precede the mediator and the mediator should precede the outcome. For a mediation analysis to be credible, the temporal ordering of the treatment, mediator, and outcome variables must support the temporal ordering of the proposed mediation model. When planning longitudinal data collection for mediation analyses it is crucial for researchers to consider optimal measurement timing, accounting for the hypothesized timing of the effects of interest. In the context of secondary data analysis, in some cases, it may be possible to conduct mediation analyses *post hoc* if the measurement timing reflects the temporal ordering of the hypothesized model. In settings where repeated waves of data were collected, it may be possible to specify a mediation model for which the baseline covariates were measured before the exposure, the exposure was measured before the mediator(s) of interest, and the mediator(s) were measured before the outcomes of interest. However, sometimes the nature of available data simply may not support analysis of the proposed mediation model. If the way the exposure, potential mediator, and outcome variables were measured does not clearly establish temporality – as is often the case in cross-sectional data – it is impossible to analytically identify temporal ordering. We also highlight that establishing temporality of covariates – relative to the exposure, mediator, and outcome – is crucial, as this differentiates baseline (pre-treatment) confounders from post-treatment confounders.

Conducting mediation analysis using cross-sectional data in which temporal ordering is not well-defined has been shown to often result in biased, misleading results (Lindenberger et al., 2011; Maxwell &

Cole, 2007; Maxwell et al., 2011; O'Laughlin et al., 2018; Selig & Preacher, 2009). Fundamentally, one of the key assumptions regarding a mediational pathway is the temporal ordering, namely the passage of time that elapses between the exposure, mediator, and outcome. As such, a mediation analysis should examine variables that are measured sequentially, rather than concurrently (O'Laughlin et al., 2018). Generally, use of cross-sectional measures of the exposure, mediator, and outcome (all assessed at the same time) would implausibly suggest that effects are instantaneous (Gollob & Reichardt, 1987). We note that, in some cases, the temporal ordering can be assumed with cross-sectional data – e.g., when using retrospective questions that refer to different time periods. However, the validity of the temporal ordering may still be questionable because of factors such as recall bias. A second key limitation of using cross-sectional data is that it prevents controlling for baseline measures of the mediator(s) and outcome(s), which has been shown to be important for reducing bias when estimating mediational effects (Selig & Preacher, 2009).

Finally, we note that temporal ordering becomes even more important – and more complex – when considering time-varying exposures, mediation processes, and/or outcomes. In such a case it is important to carefully consider – and appropriately account for – the relative temporal ordering of variables. Recent methodological developments in this area have allowed for time-varying exposures and mediators (Lin, Young, Logan, & VanderWeele, 2017; VanderWeele & Tchetgen Tchetgen, 2017), time-varying mediators and outcomes (Bind et al., 2016; Zeng et al., 2021), and time-varying mediation effects with time-varying mediators and outcomes (Cai et al., 2022; Chakraborti et al., 2022). Notably, the latter two references allow the mediated effect itself to vary as a function of time in addition to allowing values of the mediator and/or outcome to vary over time. A central challenge regarding estimation of mediational effects in the presence of time-varying exposures and mediators is time-varying confounding. Specifically, a time-varying exposure serves as a "post-treatment" confounder regarding the mediator-outcome effect (similar to **Figure 4** but replacing *A* and *X* with $A_t$ and $A_{t+1}$, respectively). As we discuss in the next section, not all mediational effects are able to be estimated in this context (i.e., see section on *interventional effects*).

**CHALLENGE 4: Selecting mediation effects that reflect the scientific question of interest**

The incorporation of a causal inference perspective in mediation analysis reveals that causal effects can be defined in different ways depending on which potential outcomes are contrasted. The well-known effect types include natural (in)direct effects (Pearl, 2001; Robins & Greenland, 1992), controlled direct effects, and interventional (in)direct effects (Lok, 2016; Lok & Bosch, 2021; Vanderweele et al., 2014; Vansteelandt & Daniel, 2017). Also, other effects can also be defined flexibly depending on the specific scientific question (Nguyen et al., 2020). The discussion of these various effects in the literature generally focuses on the issue of identification and estimation, offering little guidance to applied researchers regarding how to select which type of mediation effect is most appropriate for the research question at hand. In this section, we give a brief introduction to these effects, describing the motivation for using each effect type.

*Natural direct and indirect effects*

Natural direct and indirect effects are motivated by the desire to explain the total causal effect. They are defined to decompose the total effect, i.e., the combination of the natural direct effect and the natural indirect effect is equal to the total effect. Natural effects are popular because they speak to the original motivation for mediation analysis, namely, to explain the total causal effect through effect decomposition.

These effects are defined based on potential outcomes. Consider the case in which *A* is a binary exposure variable, indicating the exposed ($A = 1$) or the comparison ($A = 0$) condition. Each individual has a potential outcome had they received the exposure (denoted $Y_1$) <u>and</u> a potential outcome had they received the comparison condition (denoted $Y_0$). While both potential outcomes exist for all individuals, we can observe only one of the two. Formally, $Y = Y_A$, i.e., the observed outcome reveals the potential outcome corresponding to the individual's actual exposure condition; this is the *consistency* assumption. For an individual, the effect of the exposure on the outcome (which in mediation analysis is called the total effect) is taken to be the difference between these two potential outcomes, $TE = Y_1 - Y_0$. Generally, one takes the average of this difference across individuals to obtain a population average effect.

The definition of natural direct and indirect effects relies on a "nested" type of potential outcomes that involves the mediator $M$. Note that the mediator is an intermediate outcome of the exposure and itself has potential values, $M_1$ and $M_0$. Let $a$ and $a'$ be two indices that could be either 1 or 0. Let $Y_{aM_{a'}}$ denote the potential outcome for the case when the exposure is equal to $a$ and the mediator is equal to $M_{a'}$ (the potential value of the mediator under condition $a'$). Crossing both exposure conditions with both potential mediator values yields four nested potential outcomes: $Y_{1M_1}$, $Y_{0M_0}$, $Y_{1M_0}$ and $Y_{0M_1}$. The latter two of these correspond to *cross-world* conditions, namely hypothetical (i.e., unobservable) conditions in which $a \neq a'$. Natural direct and indirect effects are contrasts of these potential outcomes. There are two natural indirect effects, $NIE_1 = Y_{1M_1} - Y_{1M_0}$ and $NIE_0 = Y_{0M_1} - Y_{0M_0}$, each contrasting a change of mediator from $M_0$ to $M_1$ while holding the exposure constant. There are two natural direct effects $NDE_0 = Y_{1M_0} - Y_{0M_0}$ and $NDE_1 = Y_{1M_1} - Y_{0M_1}$, each contrasting a change of exposure from 0 to 1 while holding the mediator constant. These effects form two pairs, the $[NDE_0, NIE_1]$ pair and the $[NDE_1, NIE_0]$ pair, each of which sum to the overall total effect, $TE = Y_{1M_1} - Y_{0M_0}$.

We highlight that, when reporting natural effects, the researcher must select which decomposition(s) of the total effect to report – a choice that has generally received little attention in the literature. Notably, these two decompositions become equivalent if there is no exposure-by-mediator interaction. Methodological papers tend to either present both decompositions or to present only the decomposition into $NDE_0$ and $NIE_1$ (without discussing the motivation for this choice). Recent work by Nguyen et. al. (2020) provides suggestions on which natural effect decomposition to select. They propose using the [$NDE_0$, $NIE_1$] pair if the research question is "Is there a mediated effect?" or "Is the causal effect (partly) mediated by the proposed mediator?" Alternatively, they propose using the [$NDE_1$, $NIE_0$] pair if the question is "In addition to the mediated effect, is there a direct effect?" or "Does the exposure influence the outcome in other ways, not through this mediator?" Finally, they recommend presenting both decompositions if there is no prior assumption or preferred question regarding direct or indirect effects. We note that there are alternatives to the two decompositions discussed above that may be of interest in particular settings; see (Hong et al., 2015; Nguyen et al., 2022; VanderWeele, 2015) for more details.

Additionally, we highlight that these are conceptual, non-parametric definitions for the NDE and NIE that do not require a specific estimation approach for modeling $M$ and $Y$. Thus, after selecting the NDE and NIE as the mediation effects of interest, the applied researcher needs to choose a specific estimation approach (e.g., inverse probability weighting, g-computation). Analytic considerations regarding the estimation approach are beyond the scope of this paper – please see (Nguyen et al., 2023) for a detailed discussion.

In light of the call for "pragmatic epidemiological research" (i.e., focusing on real-life exposures or interventions applied in routine (rather than optimal) community settings), estimands that rely on hypothetical "cross-world" counterfactuals may seem impractical. However, these cross-world counterfactuals are a means to an end to obtain estimates for the NDE and NIE, which can indeed offer meaningful and pragmatic insights. In particular, mediation analyses often focus on identifying potential (actionable) mediators in health-related processes, with the intention of developing / refining interventions that target mediators.

*Interventional effects*

*Interventional effects* refer to a broad class of causal contrasts that involve hypothetical conditions where the exposure and/or mediator is intervened on and set to either a specific value or a specific distribution. Most of these effects are defined in the aggregate rather than individual levels. Since these effects are about hypothetical situations (including hypothetical modifications of exposure or hypothetical interventions on the mediator), they should be interpreted as "predictive" effects. A key difference between interventional effects and the natural (in)direct effects above is that, in principle, one could conduct an experiment contrasting the interventional effect conditions of interest, whereas natural (in)direct effects are not experimentally testable (Robins, 2003). Additionally, interventional effects do not decompose the total effect on the outcome while natural (in)direct effects do.

<u>*Controlled direct effects*</u>. The simplest effects in this class are controlled direct effects, which contrast conditions where both exposure and mediator are set to specific, known, values. A controlled direct effect (for mediator level $m$) is the causal effect of the exposure on the outcome if the mediator were held constant at level $m$ for the entire population (across both treatment conditions) (Robins & Greenland, 1992). That is, $\text{CDE}(m) = \text{E}[Y_{1m}] - \text{E}[Y_{0m}]$, where $\text{E}[\cdot]$ is notation for the population mean. (A controlled direct effect can also be defined at the individual level as $Y_{1m} - Y_{0m}$.) While this contrast compares conditions with and without the intervention, it is defined in a context where regardless of intervention condition, the mediator is fixed to a specific value. Controlled direct effects are thus only of interest in very specific situations where an anticipated/imagined external manipulation is believed to have this mediator-fixing impact.

*Illustrative example:* Nguyen *et al.* (2020) give the example of a childhood burn prevention campaign, which sought to raise parental awareness of common causes of childhood burns, including excessively high settings on household water heaters. It was determined that one mechanism of the intervention's impact on

> the rate of childhood burns (the outcome) was through reductions in water heater temperatures (the mediator). A new city ordinance will soon require that household water heaters must not be set higher than 120°F (the $m$ value). City officials want to know how much impact continuing the burn prevention campaign will have, considering the new ordinance that fixes the mediator level – this effect can be estimated by the controlled direct effect $\text{CDE}(m)$.

We note that a more generalized version, referred to as a *generalized direct effect*, holds the mediator constant at a given distribution, allowing some variation in mediator level across individuals, rather than a fixed level $m$. Also, we highlight that there is no corresponding notion of a "controlled indirect effect." This is not surprising because we can define controlled direct effects without conceptualizing mediation at all; a controlled direct effect is simply the effect of the exposure in a specific setting (where variable $M$ is controlled). Thus, estimation of controlled direct effects will only provide insight regarding direct effects, not indirect effects.

*Interventional (in)direct effects.* Historically, one of the motivations for defining interventional (in)direct effects is that these effects – unlike natural (in)direct effects – are identified in the presence of mediator-outcome confounders that are influenced by exposure (i.e., post-treatment confounders). The *interventional direct effect* (also called *randomized, stochastic, or organic direct effect*) is a specific type of generalized direct effect in which the mediator distribution is held constant to the distribution of potential mediator value $M_1$ (or $M_0$), given a set of pre-exposure covariates $C$ (Didelez et al., 2006; Lok, 2016; Vanderweele et al., 2014). The key difference between interventional direct effects and natural direct effects is that the latter involve setting the mediator to individual-specific potential mediator values $M_1$ (or $M_0$), whereas interventional effects shift the mediator distribution to be equal to a conditional distribution of the potential mediator. Also, interpretation of natural effects does not invoke the notion of a hypothetical intervention; rather, natural effects seek to decompose the observed total effect.

In order to formally define interventional (in)direct effects, we need additional notation: let $\mathcal{M}_{a|C}$ (with a script letter $\mathcal{M}$) denote the distribution of potential mediator $M_a$ given $C$ and let $Y_{a\mathcal{M}_{a'|C}}$ denote the outcome that would arise if the exposure were set to $a$ and the mediator distribution was intervened on and set to the distribution of $M_{a'}$ given $C$. The two interventional direct effects are defined as: $IDE_0 = \text{E}[Y_{1\mathcal{M}_{0|C}}] - \text{E}[Y_{0\mathcal{M}_{0|C}}]$ and $IDE_1 = \text{E}[Y_{1\mathcal{M}_{1|C}}] - \text{E}[Y_{0\mathcal{M}_{1|C}}]$. Analogously there are two interventional indirect effects: $IIE_0 = \text{E}[Y_{0\mathcal{M}_{1|C}}] - \text{E}[Y_{0\mathcal{M}_{0|C}}]$ and $IIE_1 = \text{E}[Y_{1\mathcal{M}_{1|C}}] - \text{E}[Y_{1\mathcal{M}_{0|C}}]$, each of which contrasts shifting the mediator distributions from $\mathcal{M}_{0|C}$ to $\mathcal{M}_{1|C}$, while holding the exposure fixed. Intuitively, shifting the mediator distribution is equivalent to assigning each individual a random draw from that specific distribution (Nguyen et al., 2020). Interventional (in)direct effects do not decompose the total effect, as they are not designed to do so.

In practice, researchers may choose to estimate interventional effects in settings where natural effects are not identified (i.e., the presence of post-treatment confounders) – in this sense, interventional effects might serve as an approximation for the unidentified natural (in)direct effects. However, the motivation of an analysis estimating interventional effects is generally different than that of an analysis estimating natural effects. As detailed by Nguyen et al. (2020), interventional effects seek to answer "what if" questions about what magnitude of change could be expected at the population level if a hypothetical intervention was implemented.

*Illustrative example:* Consider a high school that currently has a substance use prevention program that comprised several different components, including self-efficacy building, physical exercise, and mindfulness. Another school in the district is considering implementing this program, yet due to budget constraints they would drop the mindfulness component (here, our potential mediator *M*). An estimate of the level of substance use-related incidents (here, *Y*) this school could anticipate if they implemented this modified program (here, $A = 1$) would be provided by $IDE_0$. Here, $IDE_0$ contrasts the effect of the proposed modified program which would not impact the mindfulness distribution ($E[Y_{1\mathcal{M}_{0|C}}]$) to a hypothetical control condition (e.g., no prevention program), assuming a similar lack of impact on student mindfulness ($E[Y_{0\mathcal{M}_{0|C}}]$).

*Defining interventional effects more flexibly.* Rather than defaulting to interventional (in)direct effects, Nguyen et al. (2021) advocates for defining interventional effects flexibly to best match the scientific research question. This requires being thoughtful about (a) which active intervention condition is of interest, and given this choice, (b) which condition is the appropriate comparison condition. In the prior example of modifying the substance use prevention program, a more relevant comparison condition may be the true status quo $E[Y_0]$, without imposing the distributional assumption about mindfulness $E[Y_{0\mathcal{M}_{0|C}}]$. As such, a more meaningful interventional effect may be given by $E[Y_{1\mathcal{M}_{0|C}}] - E[Y_0]$. This more general framework for interventional effects provides flexibility to define causal effects that better accommodate a much broader range of research questions. For an example of how flexibly defining effects readily applies to research on health/social disparities, see (Nguyen et al., 2020). As such research involves imagining alternative worlds where social and structural elements that contribute to disparities were mitigated or neutralized, there is no standard pair (or set) of contrasts. Rather, each imagined alternative world calls for an appropriate contrast that can be accommodated within this flexible framework.

**CHALLENGE 5: Assessing the validity of underlying assumptions required for estimation**

This section will discuss the necessary assumptions for identification (i.e., identifying the potential outcomes, which are unobserved, from the observed data) for each of the mediational effects defined in the previous section – natural effects, interventional effects, and controlled direct effects. Importantly, all of these effects require several assumptions regarding confounding – referred to as "ignorability assumptions" or "sequential randomization assumption" – in order to be interpreted as causal. We note that our focus is on the single mediator case – the context of multiple mediators may require additional assumptions (which may vary depending on the proposed interrelationship between the mediators).

Natural effects require the most stringent assumptions. Specifically, to identify natural direct and indirect effects, four assumptions regarding confounding are needed: no unobserved confounding of the (1) exposure-mediator relationship, (2) exposure-outcome relationship, or (3) mediator-outcome relationship, as well as (4) no observed or unobserved confounders of the mediator-outcome relationship that have been influenced by the exposure (also referred to as "post-treatment confounders" or "intermediate confoundingconfounders"). These confounding assumptions are depicted in **Figure 5**.

**Figure 5.** The four types of confounding that are assumed not to exist for identification of natural direct and indirect effects

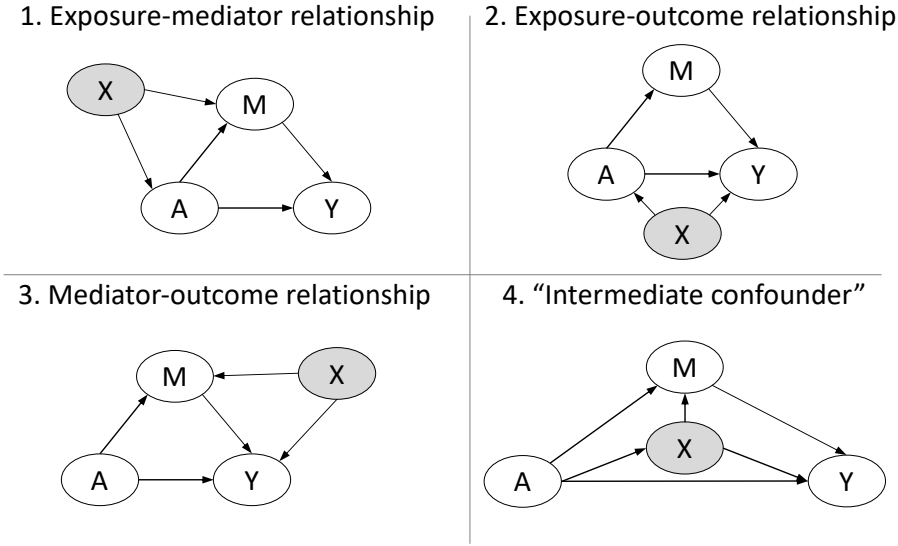

We note that certain modern estimation methods for the NIE/NDE may relax some of these assumptions in some contexts, and that certain types of potential confounding may be able to be addressed analytically through techniques such as regression adjustment or propensity score weighting (see (Nguyen et al., 2022) for more details). If there is random assignment to the intervention or exposure variable then the assumptions of no unmeasured confounders of the exposure-mediator and exposure-outcome relationships should hold; however, the other two assumptions regarding confounding of the mediator-outcome relationship are not guaranteed to hold even when there is random assignment to the exposure variable. The no intermediate confounders assumption is particularly stringent in that it means there are no time-varying (i.e., post-treatment) confounders and that any additional mediators must be conditionally independent of the mediator of interest, yet this assumption was traditionally required to identify mediation effects, starting with Baron and Kenny (1986), Robins and Greenland (1992), and Pearl (2001). Subsequent work identified approaches and required assumptions to assess mediation in the presence of an exposure-mediator interaction, including (Preacher et al., 2007), (Vansteelandt & Vanderweele, 2012), (Coffman & Zhong, 2012), (Valeri & Vanderweele, 2013), and (Tchetgen Tchetgen & Vanderweele, 2014). The development of interventional effects by (Vanderweele et al., 2014) was motivated by a desire to identify causal mediation effects that do not require the assumption of no intermediate confounding; this approach was later extended to allow for multiple mediators by (Vansteelandt & Daniel, 2017).

Interventional effects require fewer assumptions to identify than the natural effects; yet they do not decompose the total effect like the natural direct and indirect effects do. Identification of the interventional direct and indirect effects requires the assumption of no unobserved confounding of the (1) exposure-mediator relationship, (2) exposure-outcome relationship, or (3) mediator-outcome relationship. If, additionally, there are no observed or unobserved confounders of the mediator-outcome relationship that have been influenced by the exposure, then the interventional direct and indirect effects are equal to the natural direct and indirect effects (VanderWeele & Tchetgen Tchetgen, 2017).

Identification of the controlled direct effect requires the fewest confounding assumptions, namely that there are no unmeasured confounders of the (1) exposure-outcome relationship and (2) mediator-outcome relationship. If there is random assignment to the exposure, then the former should be satisfied, although it is still a concern in observational studies. As randomization to the exposure does not imply randomization to levels of the mediator, the second confounding assumption will be relevant in both randomized and observational studies. Although it is easier to identify the controlled direct effect, it unfortunately does not have a corresponding indirect effect as mentioned in the previous section.

Additionally, identification of natural effects, interventional effects, and controlled direct effects all also require the assumption of positivity, that is, that individuals have some positive probability of receiving each level of the treatment and each level of the mediator. See (Nguyen et al., 2022) for further discussion of the positivity assumption for each effect type.

Table 2. Summary of required assumptions regarding unobserved confounding and positivity required for estimation of natural effects, interventional effects, and controlled direct effects

|  | No *unobserved* confounders of the: | | | No *observed* or *unobserved* confounders of the: mediator-outcome relationship that have been influenced by the exposure | Positivity |
|---|---|---|---|---|---|
|  | exposure-mediator relationship | exposure-outcome relationship | mediator-outcome relationship | | |
| Natural effects | ✓ | ✓ | ✓ | ✓ | ✓ |
| Interventional effects | ✓ | ✓ | ✓ | (if true, equivalent to natural effects) | ✓ |
| Controlled direct effect |  | ✓ | ✓ |  | ✓ |

Notably, the previously described assumptions regarding no unmeasured confounding cannot be empirically assessed, as they fundamentally relate to unobserved variables. Rather, sensitivity analyses have been developed that assess the potential bias that would arise from different types of unmeasured confounding. These sensitivity analyses can be used to determine the magnitude of unobserved confounding that would qualitatively change the results of a given mediation analysis. The specific choice of sensitivity analysis depends on which mediation effects are being estimated. For the natural direct and indirect effects, Imai et. al. proposed a sensitivity analysis to the no-unmeasured-confounding assumptions (Imai et al., 2010) and for the presence of a post-treatment confounder (Imai & Yamamoto, 2013) and implemented them in the *mediation* R package. VanderWeele proposed a sensitivity analysis for the no-unmeasured-confounding assumptions used in identifying controlled direct effects (VanderWeele, 2010) as well as sensitivity analysis methods for natural direct and indirect effects regarding mediator-outcome confounders affected by the exposure (i.e., post-treatment confounders) (VanderWeele & Chiba, 2014). VanderWeele's sensitivity analyses are implemented in the *CMAverse* R package (Shi et al., 2021). Additional work has proposed sensitivity analyses for natural effects to simultaneously assess unmeasured confounding in the mediator-outcome, exposure-outcome, and exposure-mediator relationships using a Bayesian approach (McCandless & Somers, 2019). Hong et. al. introduced a weighting-based sensitivity analysis (an extension of the ratio-of-mediator-probability weighting (RMPW) method) to assess potential bias arising from omitted pre-treatment or post-treatment covariates; this method has minimal functional form assumptions and accommodates a broad range measurement scales for the mediator, outcome, and omitted covariates (Hong et al., 2018). Additional sensitivity analysis strategies include: using multiple measures of the same mediating construct to bolster evidence; conducting falsification tests with variables hypothesized to not be mediators; or assessing robustness of findings to the choice of different estimation methods (Coffman et al., 2016).

More recently, more comprehensive approaches to sensitivity tests have been proposed, referred to as "multiverse approach" or a "multi-model analysis" (Steegen et al., 2016; Young & Holsteen, 2017). These approaches seek to quantify how robust findings are to *all* decisions and specifications made throughout the data analytic processes (Steegen et al., 2016). Recent work has detailed how to take a multiverse approach in the context of mediation analysis (Rijnhart et al., 2022). Specifically, the authors highlight that mediation analysis, by virtue of the addition of an intermediate mediation variable, entail more analytic decisions regarding operationalization of variables and model specification compared to bivariate (i.e., non-mediation) models that simply regress an outcome on an exposure. As such, the multiverse examined in this type of sensitivity analysis will be larger and more complex than multiverses corresponding to bivariate models (Rijnhart et al., 2022). Multiverse methods can generate specification curves that quantify the impact of analytical decisions on the magnitude and statistical significance of the direct, indirect, and total effect estimates (Rijnhart et al., 2022). Overall, we note that most of the sensitivity analyses described above are for the context of a single mediator – to date, there are few sensitivity analysis methods for the context of multiple mediators or time-varying mediation, although this is an area of ongoing methodological work.

In addition to identification assumptions specific to mediation, mediation analyses are also subject to the standard assumptions related to the specific statistical modeling approach used. For example, in the context of generalized linear models, the choice of a log function confers specific assumptions – e.g., a logistic regression model requires a rare outcome assumption, which is not necessary if using a log-linear (i.e., Poisson) model (Valeri & Vanderweele, 2013). Unlike identification assumptions, it may be possible to empirically assess the plausibility of modeling assumptions (e.g., rare outcome assumption can be checked by examining the prevalence of the outcome in the data). We highlight that identification and modeling assumptions can become intertwined in the context of mediation analysis. For example, one way around the assumption of no post-treatment confounders of the mediator and outcome is to assume that there are no interactions between the treatment and the mediator; this parametric modeling assumption, when combined with an assumption of linear relationships among all the variables, serves to identify the direct and indirect effects in the traditional mediation model. Additionally, if there is no exposure-mediator interaction, the controlled direct effect of the treatment will be equal to the natural direct effect. The analyst may then employ IPTW to adjust for post-treatment confounders.

**CHALLENGE 6: Addressing measurement error regarding the mediator**

While potential bias arising from violations of the assumptions related to confounding (discussed in the prior section) has received much attention in the mediation literature, potential bias arising due to measurement error has received much less discussion to date. To ensure plausibility of the underlying consistency assumption in mediation analysis, it is imperative that the variables used in the analysis – i.e., treatment variable, mediator(s), outcome(s), and other covariates -- are measured without error. In many fields of study, however, mediators are likely measured with error. For example, psychological variables are not observed directly but measured imprecisely through multiple indicators. Educational tests provide indirect estimates of the latent abilities of interest, as do survey measures of attitudes and beliefs. Self-reports of substance use and other behaviors are also known to be measured with error; for examples see: (Cole & Preacher, 2014; Del Boca & Noll, 2000; Johnson & Fendrich, 2005; Morral et al., 2000; Natarajan et al., 2010).

> *Illustrative example:* Consider again the high school substance use prevention program comprised of self-efficacy building, physical exercise, and mindfulness components. As part of the program evaluation, students completed surveys with scales measuring their mindfulness, measured with multiple items or indicator variables. The scores on the mindfulness scales contain measurement error and using them in mediation analysis would result in bias. For example, if the reliability of the scale score is 0.80 (often considered an acceptable level foror reliability) and the models for the mediator and outcome are linear, then using the formula for the bias in Valeri *et al.* (2014) the NIE would be biased toward zero by about 20 percent, i.e., a positive NIE would be 20% too small and a negative NIE would be too large by 20%. Such bias could lead to inaccurate conclusions about the importance of mindfulness in the substance use prevention program. Analytic methods that account for measurement error should be used.

Much work regarding mediation measurement error has been in the context of the regression-based approach to mediation analysis, in which one estimates parameters from both an outcome and mediator regression model, and then calculates direct and indirect causal effects as functions of those regression parameters (Valeri & Vanderweele, 2014). It is well-known that measurement errors can create bias in regression analysis (Carroll et al., 2006; Fuller, 1987). As early as 1981, (C.M. Judd & D.A. Kenny, 1981) discussed that mediator measurement error has the potential to bias estimates of the mediated effects. Baron and Kenny (1986) postulated that because measurement error in predictor variables in linear regression tends to attenuate the regression coefficients, measurement error in a mediator would be expected to bias the coefficient on the mediator in the outcome model toward zero (i.e., leading to indirect effects biased toward zero). Given the additive relationship between the direct and indirect effect, mediator measurement error would also be expected to bias the coefficient for the treatment in the outcome model, leading to direct effects biased away from zero.

One method of addressing measurement error in this context is to apply a correction factor, based on the theorized magnitude of the bias. Ogburn and VanderWeele show that, when a single dichotomous mediator is measured with error, (1) the NIE of the error-prone mediator will generally be between zero and the true NIE for the error-free mediator and (2) the NDE will be between the true NDE and the total effect (Ogburn & VanderWeele, 2012). These results are nonparametric and require no assumptions of a linear or generalized linear model for the outcome or mediator. In the context of linear, logistic regression, and generalized linear models for the outcome and mediator, numerous authors have given formulas for the bias caused by measurement errors in dichotomous and continuous mediators as well as the bias in the direct and indirect effects (Fritz et al., 2016; le Cessie et al., 2012; Valeri et al., 2014; Valeri & Vanderweele, 2014; VanderWeele et al., 2012). Specifically, le Cessie et al. (2012) only considers direct effects; (VanderWeele et al., 2012) suggest how to extend the results to indirect effects. (Valeri et al., 2014; Valeri & Vanderweele, 2014) further extend this work by developing correction approaches that allow for exposure–mediator interaction for both binary and continuous mediators.

An alternative approach to addressing measurement error in mediation analysis is use of structural equation models (SEM), which have been used for mediation analysis when there are multiple (error-prone) indicators for the latent variable for over 30 years (Brown, 1997; C. M. Judd & D. A. Kenny, 1981). For example, depression can be considered a latent construct, in that is not directly measured, but rather typically assessed via responses to multiple questions that are all related to depression. Using a SEM, one can specify a measurement model linking the multiple indicators to a latent variable and then specify models for the exposure, latent variables, and outcomes to study the indirect effects of an exposure through a latent variable such as depression (Hayes & Preacher, 2010).

Alternatively, when the reliability of the error-prone measure is known, methods for directly correcting measurement error have been proposed, including the simulation and extrapolation method (SIMEX) (Cook & Stefanski, 1994) or regression calibration (Carroll et al., 2006). In some fields, information about the measurement error may be available – e.g., in the context of educational or psychological measurement, in which psychometric properties of tests are explicitly measured. In other applications, such as self-reported substance use, the distribution of errors is generally unknown, although widely considered to exist. In these cases, direct adjustment will be impossible unless information about the measurement errors can be obtained.

Finally, as highlighted by (Ledgerwood & Shrout, 2011), in some contexts it may be more desirable to use the biased estimates rather than correcting for measurement error, even when corrections are possible. They note that unadjusted estimated indirect effects using the observed error-prone mediator variables may be more precise than those obtained using latent regression or SEM approaches. Given that statistical power for estimation of indirect effects may be limited in some contexts, a more precise, but biased, estimate may provide more powerful hypothesis tests regarding the direct and indirect effects compared to a bias-corrected estimate. Overall, whether correcting for measurement error is desirable will depend on the details of any application, yet it is important that standard errors be calculated appropriately, and the relative precision be considered when correcting for measurement error in mediation analysis.

**CHALLENGE 7: Clearly reporting results from mediation analyses**

As noted earlier, mediation analyses have become increasingly common in recent years and some recent NIH funding mechanisms require investigation of mediating mechanisms. However, as we have highlighted in Challenges 1-6, applied researchers face numerous methodological challenges when conducting mediation analyses. This is reflected in the literature: multiple methodological reviews of the applied mediation literature have found that the execution and presentation of applied mediation studies is of varying quality and often lacking (e.g., (Cashin et al., 2020; Gelfand et al., 2009; Liu et al., 2016; Rijnhart, Lamp, et al., 2021; Stuart et al., 2022; Vo et al., 2020)). For example, while (Rijnhart, Lamp, et al., 2021) found that approximately 75% of the 174 mediation studies they reviewed included a figure showing the proposed mediation model, Cashin et al. (2020) found that less than 15% of studies reviewed actually presented a conceptual justification for examining the proposed mediation pathway. Many studies performed mediation analyses using variables with murky temporal ordering -- (Vo et al., 2020) found that the mediator and outcome were measured simultaneously in over 50% of mediation studies they reviewed and (Rijnhart, Lamp, et al., 2021) found that nearly half of the studies reviewed used a cross-sectional design. Focusing on the psychology and psychiatry literature, Stuart et al. (2022) found that only about 25% of the 206 mediation analyses reviewed had full temporal ordering of exposure, mediator, and outcome, and that fewer than half of the papers controlled for confounders. Currently, the majority of mediation studies use a traditional mediation approach (e.g., Barron & Kenny), with approximately 5-15% of studies using a causal mediation framework (Rijnhart, Lamp, et al., 2021; Stuart et al., 2022). Both (Stuart et al., 2022; Vo et al., 2020) found that a minority of studies stated the underlying assumptions of their mediation analysis and both (Rijnhart, Lamp, et al., 2021; Stuart et al., 2022) found that less than 5% of studies performed any type of sensitivity analysis.

In order to advance the state of the science regarding mediation, reporting guidelines – i.e., AGReMA or "<u>A G</u>uideline for <u>Re</u>porting <u>M</u>ediation <u>A</u>nalyses" – have recently been proposed by Cashin et al (2020) and (Lee et al., 2021). While reporting guidelines have existed for randomized trial study designs (see Consolidated Standards of Reporting Trials (CONSORT)) and for observational studies (see

Strengthening the Reporting of Observational Studies in Epidemiology (STROBE)), formal guidance regarding mediation analyses was previously quite limited. AGReMA was developed via a systematic, expert consensus process that focused on identifying essential elements that should be reported in mediation studies using experimental or observational data. AGReMA emphasizes that studies should clearly state and justify the causal question(s) being investigated, as highlighted by the following recommendations: (1) clearly state the mediation question of interest; (2) provide rationale for investigating the specified mediation pathway(s); (3) present a figure depicting the underlying causal model (including confounders); and (4) state which specific mediation effects are of interest. Additionally, AGReMA highlights key recommendations regarding measurement and analysis, including: (1) specify how and when all variables (e.g., exposures, mediators, outcomes, confounders) – were measured; (2) report details of the statistical methods used to estimate causal effects, including how potential confounding was addressed; and (3) report and interpret estimated effects, (including uncertainty estimates). AGReMA also recommends that studies: (1) specify all assumptions underlying the analyses reported; (2) discuss the plausibility of underlying assumptions; (3) report results from sensitivity analyses assessing the robustness of study findings; and (4) discuss study limitations, additional potential sources of bias, and generalizability of findings. There is both a long-form AGReMA checklist for studies that are conducting mediation as a primary analysis, as well as a short-form checklist for when mediation is a secondary study aim (Lee et al., 2021).

We highlight that AGReMA provides guidance on *reporting* of mediation analyses, rather than guidelines regarding the *analytic execution* of mediation analyses. By encouraging greater transparency and more detailed methodological description when reporting mediation studies, the AGReMA guidelines seek to make it easier to assess the methodological quality of mediation studies. While these guidelines provide a comprehensive overview of best reporting practices, the ultimate success of these endeavors ultimately lies with journal editors, reviewers, and granting agencies, and researchers themselves to ensure that standards are met in applied studies.

In additional to AGReMA, other guidance for reporting mediation analyses has been offered in the literature. For example, concerns have been raised about the utility of an often-reported mediation quantity, the "proportion mediated" which is generally calculated as the ratio between the NIE and total effect. Specifically, this quantity may be ill-defined in the presence of exposure-mediator interactions (i.e., the proportion mediated may vary across levels of the mediator) or "inconsistent mediation" when the total effect is near zero or of the opposite sign as the NIE (VanderWeele, 2013). VanderWeele alternatively proposes the reporting of the proportion of the total effect that could be eliminated by an intervention that set the mediator to a fixed level (VanderWeele, 2013).

Finally, conducting systematic reviews and meta-analyses of mediation studies is an emerging area, given the growing number of published mediation studies. However, as we have highlighted in this paper, any single mediation analysis entails a myriad of decision points regarding variable measurement, hypothesized causal structure, mediation effects of interest, and estimation methods, including strategies to mitigate potential confounding and assess robustness to violations of underlying assumptions or model misspecification. In practice, mediation analyses investigating related mediating mechanisms may vary so substantially that it is infeasible to pool any information across studies. A recent paper outlined some key difficulties regarding synthesizing mediation analyses for systematic reviews or meta-analyses falling under 3 broad challenges: (1) Identification of eligible studies; (2) Assessing the presence of different sources of biases in eligible studies, and (3) Synthesizing quantitative findings across eligible studies (Vo & Vansteelandt, 2022). Ongoing methodological work is needed in this area.

CONCLUSION

This paper provides a summary of 7 key challenges encountered when conducting mediation analysis as well as a discussion of how applied researchers can address each challenge. Mediation analysis entails an examination of causal mechanisms – this fundamental motivation should be reflected in the design and interpretation of mediation analyses. When designing mediation studies, it is important to clearly identify the objectives of mediation analysis and how these differ from objectives of other types of "third variable" analyses. Additionally, in the literature, more attention is needed regarding the assumptions inherently underlying mediation analysis. The field of causal mediation has much to offer in this regard, as the application of a causal inference framework has served both to differentiate specific types of mediation effects that can be estimated as well as to formalize the necessary assumptions underlying each effect type. Mediation analyses generally require more assumptions than traditional bivariate exposure-outcome analyses, underscoring the importance of assessing the robustness of study findings to assumption violations via sensitivity analyses. To promote adoption of mediation sensitivity analyses, methodologists should strive to disseminate both knowledge of and accessible software to implement these methods. Finally, there is a need for better reporting of mediation analysis – the development of AGReMA represents a very promising direction for the field. Overall, we hope that this paper will help advance the practice of mediation analysis, leading to publication of more robust mediation studies in the health services and health policy fields.